\documentclass[11pt]{article}

\begin{document}

\title{Boundary layers in turbulent Rayleigh-B\'enard convection in air}

\author{Ronald du Puits, Johannes Rilk, Christian Resagk, and Andr\'e Thess \\
\\\vspace{6pt} Institute of Thermodynamics and Fluid Mechanics,
\\ Ilmenau University of Technology, 98693 Ilmenau, Germany}

\maketitle


\begin{abstract}
The boundary layer flow in a Rayleigh-B\'enard convection cell of rectangular shape has been visualized in this fluid dynamics video. The experiment has been undertaken in air at a Rayleigh number $Ra=1.3\times 10^{10}$ and a Prandtl number $Pr=0.7$. Various sequences captured at selected positions of the heating plate show that the boundary layer is a very transient flow region characterized by coherent structures that permanently evolve. It becomes fully turbulent in the areas where the large-scale circulation impinge or leave the bottom plate.
\end{abstract}

\vspace{1cm}
Whenever predicting the convective heat flux from a hot or cold surface to a surrounding fluid assumptions on the general structure of the convective boundary layer - this is the small fluid layer adjacent to the wall - have to be made. Although this layer has been, and still is, the object of numerous experimental and numerical investigations a comprehensive model being applicable in a wide range of parameters is still missing. Our flow visualization may contribute for a better understanding of this very complex flow field particularly in gases with Prandtl number $Pr\approx0.7$.

The flow visualization was accomplished in a large-scale Rayleigh-B\'enard experiment, the ''Barrel of Ilmenau``. This facility consists of an adiabatic cylinder with a diameter of $D=7.15$~m and a variable distance $H$ between the heated bottom and the cooled top plate adjustable from $H=0.05$~m and $H=6.30$~m. It is filled with ambient air with Prandtl number $Pr=0.7$. In order to confine the convective flow in a single plane a rectangular box with the size of $2.5~\times~2.5~\times~0.6~\rm{m^3}$ (height $\times$ width $\times$ depth) has been separated from the entire space of the test cell. A temperature difference $\Delta T=10~\rm{K}$ has been applied between the heating and the cooling plates resulting in a Rayleigh number $Ra=1.3\times10^{10}$. A vertical Laser light sheet, as high as 70~mm and as thick as about 2~mm, has been generated along the entire length of the bottom plate. Cold-atomized droplets of Di-ethyl-hexyl-sebacat (DEHS) of $1~\rm{\mu m}$ in size have been added to the air. Using a CANON EOS 600D camera in the High-Definition video mode ($1920\times1080$~pixels, 30~frames per second) movies of the particle motion adjacent to the heating plate have been captured at various positions along the path of the convective circulation.

Those sequences that have been embedded in the movie show the flow field at the center position of the plate as well as at two specific areas where the large-scale circulation impinges or leaves the plate. In the center position the flow field is characterized by a very transient behavior. Coherent structures like thermal plumes or eddies permanently evolve. They always penetrate the boundary layer which is as thick as about one third of the total height of the visualized flow field. Only from time to time the flow becomes laminar as shown right at the beginning and at the end of the first sequence. At the other two areas close to the vertical side walls the boundary layers behave virtually turbulent over the entire duration of the visualization.

Our movie clearly demonstrates that the boundary layer in turbulent Rayleigh-B\'enard convection is a very transient flow region even at the relatively moderate Rayleigh number $Ra=1.3\times10^{10}$. It is strongly characterized by coherent structures breaking again and again the laminar structure of the flow along the surface of the plate. At those areas where the large-scale circulation impinge or leave the bottom plate the flow field tends to become fully turbulent.

We wish to acknowledge the financial support of the Deutsche Forschungsgemeinschaft (PU436-1), the Thueringer Ministerium fuer Wissenschaft, Forschung und Kunst and the European Union (COST MP0806)
for the work reported in this paper. Furthermore we thank Robert Kaiser and Vigimantas Mitschunas for their technical assistance.



\end{document}